\documentclass{article}
\usepackage[multiple]{footmisc}
\usepackage[T1]{fontenc}
\usepackage[margin=1in]{geometry}
\usepackage{graphicx}
\usepackage{bm}
\usepackage{authblk}
\usepackage{amsmath,amsfonts,amsgen,amsbsy,amsopn,amstext,amssymb}
\usepackage{epstopdf}
\usepackage{xcolor}
\pdfoutput=1

\begin{document}

\title{Higher-order mode filtering by a resistive layer}
\author{Svetlana Kuznetsova$^{1,}$ \footnote{Svetlana.Kuznetsova@univ-lemans.fr} }
\footnotetext[1]{Current affiliation: CNRS, Centrale Lille, ISEN, Univ. Lille, Univ. Valenciennes, UMR 8520 - IEMN, Avenue Henri Poincare, Villeneuve-D’Ascq, 60069, France}
\author{Yves Aur\'egan\footnote{yves.auregan@univ-lemans.fr}}
\author{Vincent Pagneux\footnote{vincent.pagneux@univ-lemans.fr}}
\affil{Laboratoire d’Acoustique de l’Universit\'e du Mans (LAUM),
UMR 6613, Institut d’Acoustique - Graduate School (IA-GS),
CNRS, Avenue O. Messiaen, F-72085 Le Mans Cedex 9, France}

\date{\today} 

\maketitle

A method of higher-order modes filtering in an air-filled waveguide using a resistive layer is proposed. An analogue of Cremer's criterion is discussed and used to obtain the optimal modal attenuation of the non-planar waves while the plane wave is preserved. Numerical validation of the concept is performed for a straight waveguide and an abrupt expansion in a waveguide. 

\section{\label{sec:1} Introduction}
Attenuation in acoustic waveguides with lossy wall impedance has been widely studied for a long time as it has important applications for noise attenuation in ducts \cite{morse1939transmission}. 
Cremer impedance is a well known solution to try to optimise the damping of mode propagation in a duct. This impedance corresponds to the creation of an exceptional point where two modes merge  \cite{cremer1953theory,tester1973optimization,bi2015new,zhang2019cremer,aabom2021comment,lawrie2022analytic,perrey2022mode}. These modes correspond to the two lowest order modes (mode 0 and 1) and it is conjectured that this impedance leads to the highest attenuation for all propagating modes.  

On the other hand, more recently, dissipative screens have been used to filter the modes. For example, radial screens in a circular duct have been introduced to stop the spinning modes \cite{sack2016modal}. This idea has also been applied for pressure field symmetrization with a wiremesh installed in the middle of a cavity \cite{farooqui2022ultrathin}. In this case, the symmetrical modes are insensitive to the wiremesh, while all non-symmetrical modes are absorbed.

In this letter, we present a method for optimally filtering all modes except the plane mode in a two-dimensional duct with a longitudinal resistive screen.
We find an exceptional point where mode 1 and mode 2 merge: it corresponds to a critical position $d_c$ equal to 0.22659 times the height of the waveguide and a critical resistance that depends linearly on frequency.  Then, we use an analogue of Cremer's criterion to find the optimal parameters of the screen since the mode 1 and 2 are optimally absorbed at the exceptional point while mode 0 is unaffected by the wiremesh. We illustrate ability of this system to filter an acoustic field over two configurations.

\section{Modes of the waveguide with an embedded wiremesh}
\subsection{Dispersion relation for the transverse modes}
\begin{figure}[h]
 \includegraphics[width=0.5\columnwidth]{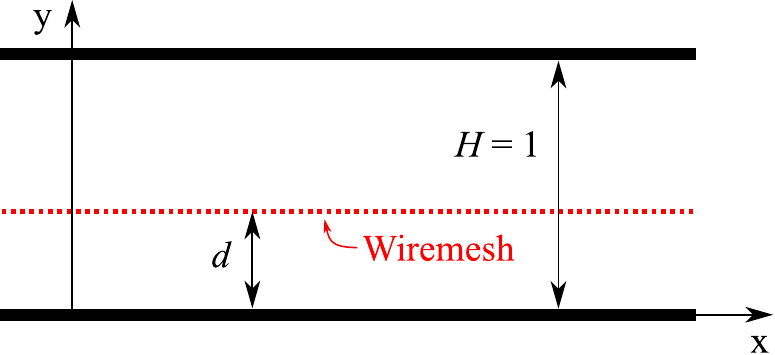}
 \caption{An infinite planar waveguide with the inserted wiremesh (red dotted line).}\label{fig:fig1}
\end{figure}

We consider the propagation of sound waves in a two-dimensional waveguide  (see Fig.~\ref{fig:fig1}) in the harmonic regime with the convention $e^{-i\omega t}$. In the following we work with dimensionless quantities, all lengths being in unit of the height of the waveguide $H$. 
In the waveguide, the acoustic pressure field obeys the wave equation 
\begin{align}
    \Delta p + k^2 p = 0,
\end{align}
where $p$ is the pressure and $k = \omega H /c_0$ is the dimensionless frequency (speed of sound $c_0$). At the wiremesh located at $y = d$ we have the pressure jump given by
\begin{align}
    [p] =  i \frac{R}{k}\frac{\partial p}{\partial y}
\end{align}
with $R$ the dimensionless (in unit of $\rho_0 c_0$, $\rho_0$ being the density) resistance of the wiremesh. The rigid walls imply $\frac{\partial p}{\partial y} = 0$ at $y = 0$ and $y = 1$.

As classically, we look for solutions of the Helmholtz equation in transverse mode form $p(x,y)=p(y) \mathrm{e}^{-i \beta x}$, we obtain:
\begin{align}
p^{\prime \prime} + \alpha^2 p=0, \,\, \mbox{ with } \quad  \alpha^2 =k^2-\beta^2,
\label{propagation}
\end{align}
 where the prime indicates the $y$-derivative. 

The boundary conditions on the walls ($y=0$, $y=1$) are $p'(0) =0$ and $p'(1) =0$. The wiremesh of resistance $R$ induces a pressure jump at $y=d$ given by $p(d^+) - p(d^-) = i C p'(d)$ with $C = R/k$, while the transverse velocity is constant $p'(d^+) =  p'(d^-)$.
Using these conditions, the dispersion relation for the transverse modes can be written as
\begin{align}
\mathcal{D} \equiv \sin(\alpha)+iC \alpha \sin(\alpha d)\sin(\alpha(d-1))=0.
\label{disp}
\end{align}
The solutions of this dispersion relation define a set of modes having complex transverse wavenumbers $\{\alpha_n\}_{n \ge 0}$. With the exception of $\alpha_0 = 0$, which corresponds to the plane wave not affected by the wiremesh since it has zero vertical velocity ($\partial p/\partial y = 0$), the $\alpha_n$ (${n \ge 1}$) follow specific paths in the complex plane as the resistance increases from $0$ to infinity. For small $C \to 0$ one has $\alpha_n \simeq \pi n+iC \pi n(\cos{2\pi nd}-1)/2$. As $C \to \infty$ a wiremesh becomes an impenetrable wall and divides the waveguide into two sections with non-interacting sets of modes  with wavenumbers $\alpha_n d \simeq \pi n - i/ C\pi n$ and $\alpha_n(1-d) \simeq \pi n - i/ C\pi n$ correspondingly.

 \subsection{Exceptional points and Cremer criterion analogue}

We now follow the same idea as Cremer \cite{cremer1953theory} to try to find the maximum attenuation of the higher order modes by the resistive wiremesh: this is why we look for the creation of the exceptional point where the first two higher transverse wavenumbers ($\alpha_1$ and $\alpha_2$) merge.  In order to find an exceptional point (double roots of $\mathcal{D}$), we impose $d \mathcal{D}/d \alpha = 0$ which leads to 
\begin{align}
\frac{d \mathcal{D}}{d \alpha}=\cos(\alpha)+ iC \left\lbrack \sin(\alpha d)\sin(\alpha(d-1)) 
+\alpha d \sin(\alpha(2d-1)) - \alpha \sin(\alpha d)\cos(\alpha(d-1))\right\rbrack=0.
\label{deriv_disp}
\end{align}
Eliminating $ iC $ between Eqs.~(\ref{disp}) and (\ref{deriv_disp}), for $\alpha \neq 0$, leads to:
\begin{align}
-\alpha \sin(\alpha d)^2 +\sin(\alpha)  \sin(\alpha d)\sin(\alpha(d-1)) + \alpha d \sin(\alpha)  \sin(\alpha(2d-1)) =0.
\label{comp}
\end{align}
For each value of $d$, the Eq.~(\ref{comp}) has several discrete solutions  $\alpha$ in the complex plane. 
Lets focus on the solution of Eq.~(\ref{comp}) $\alpha_c$ closest to 0. 
We still have to make sense of this solution and to impose
that C is real ($C=R/k$ and both $R$ and $k$ are assumed to be real)); indeed
using Eq.~(\ref{disp}) with this $\alpha_c$ we obtain a complex-valued function $C(d)$. 
Among the values of $C(d)$ only one is purely real 
\begin{align}
    C_c = 0.47606
\end{align}
for the particular value 
\begin{align}
    d_c = 0.22659.
\end{align}
Then, in the complex plane the exceptional point occurs at
 $\alpha_c/\pi = 1.13208 - 0.34857 i$, where the transverse wavenumbers $\alpha_1$ and $\alpha_2$, solutions of Eq.~(\ref{disp}), coalesce. By analogy with the Cremer criterion, we expect to obtain the strongest attenuation of the modes at this point. 
 We can notice that the resistance 
 \begin{align}
     R_c = C_c k
 \end{align}
 which allows to obtain an exceptional point is thus linearly proportional to the frequency $k$.

\begin{figure}[h]
\includegraphics[width=\columnwidth]{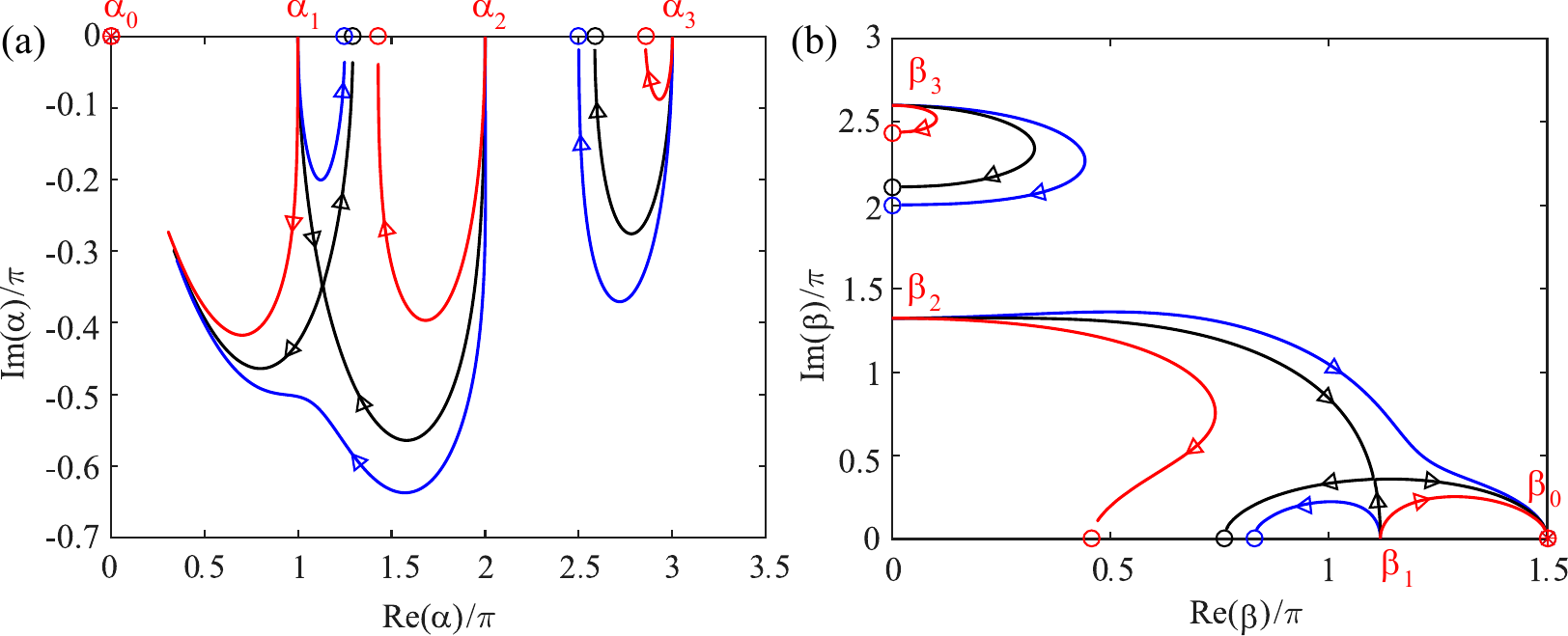}
\caption{Trajectories of the complex transverse wavenumbers $\alpha$ (a) and longitudinal wavenumbers $\beta = \sqrt{k^2-\alpha^2}$ (with positive real and imaginary parts) (b) when $C$ varies from $0$ to $2.8$. 
 $k = 1.5 \pi$ and $d = 0.2$ (blue), $d = 0.3$ (red) and $d = d_c$ (black) with the exceptional point. The circles denote the values for $C \to \infty$ and the asterisk corresponds to the mode $0$.
}\label{fig:fig2}
\end{figure}

In Fig.~\ref{fig:fig2} we plot the transverse $\alpha_n$ and longitudinal $\beta_n$ wavenumbers in the complex plane. It can be remarked that for a given $d$, the curves in the $\alpha$-plane, Fig.~\ref{fig:fig2}(a), are independent of the frequency $k$. Those curves are plotted for three values of the wiremesh position $d = 0.2$ (blue), $d = 0.3$ (red) and $d = d_c$ (black), when $C$ varies from $0$ to $2.8$. 
By choosing $k = 1.5 \pi /H$ in Fig.~\ref{fig:fig2}(b), we ensure that only two modes propagate in the empty waveguide ($C = 0$).  The plane mode ($\alpha = 0$, $\beta = 1.5 \pi$) is not affected by the wiremesh due to the zero transverse velocity. 
When $d=d_c$, $\beta_2$ (evanescent for $C=0$) coalesces  with the second initially propagating
mode $\beta_1$. 
These two modes then tend to become propagating when $C \to \infty$, one tending towards the plane mode and the other being oscillating in the transverse direction ($\lambda/2$ in the largest width).

\section{Mode filtering}

\begin{figure}[h]
\includegraphics[width=\columnwidth]{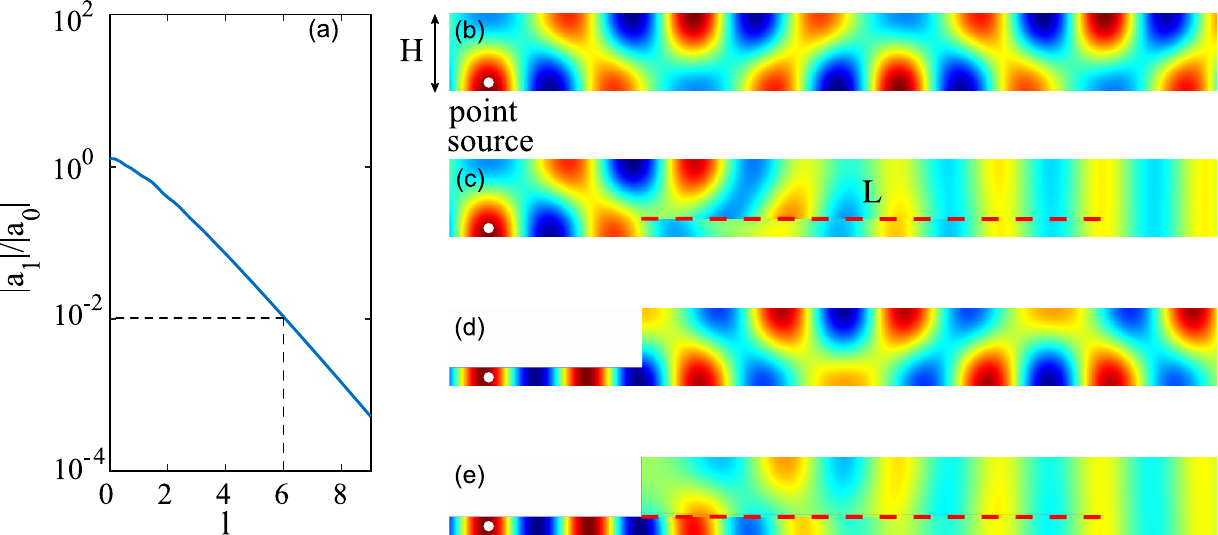}
\caption{(a) The ratio of the amplitudes on the mode $0$ and mode $1$ as a function of the wiremesh length $L$ for $d = d_c$, $R = C_c k$, (b) the pressure field in the waveguide excited by a monopole point source with  $k = 1.5 \pi$ for $R = C_c k$, $L = 6$ without a wiremesh and (c) with a wiremesh. The pressure field inside a waveguide with an abrupt expansion from $d_c$ to $1$ in the absence (d) and in the presence (e) of the wiremesh with $R = C_c k$ and $L = 6$.} \label{fig:fig3}
\end{figure}

To illustrate higher-order modes filtering, we first simulate the sound propagation in an infinite waveguide for a frequency $k=1.5 \pi/H$ for which two modes are propagating in the empty guide. In the sense of the Cremer criterion, an optimal wiremesh ($C =C_c$) of length $L=6$ is placed at an optimal distance from a wall ($d=d_c$). 
This waveguide is excited by a monopole point source located to the left of the wiremesh. 
The pressure field, computed with the Comsol Multiphysics software using perfectly matched layers on both sides to avoid reflections, is shown in Fig.~\ref{fig:fig3}(b) without a wiremesh and in Fig.~\ref{fig:fig3}(c) with a wiremesh, and demonstrates very clearly this filtering effect which allows only the plane mode to propagate. The position of the point source is illustrated by the white circle and the dashed red line represents the wiremesh.    
To quantify this filtering effect, we computed the ratio between the amplitude of the second propagating mode $|a_1|$ and the amplitude of the plane mode $|a_0|$ downstream of the wiremesh. 
The dependence of this ratio on the length of the wiremesh $L$ is shown in Fig.~\ref{fig:fig3}(a) in logarithmic scale, hence in practice one may pick the desired level of the attenuation of the first mode and adjust the length of the wiremesh in the setup. The dashed lines correspond to the parameters of Fig.~\ref{fig:fig3}(c).

 A second illustration of this filtering effect is shown in Figs.~\ref{fig:fig3}(d), (e) again for $k=1.5 \pi/H$. 
This is the case of a sudden expansion where the height of the guide changes abruptly from $d_c$ to 1. In the small guide only the plane mode propagates. 
When there is no wiremesh, Fig.~\ref{fig:fig3}(d), due to the asymmetry of this expansion both propagating modes are present in the large guide. 
By inserting the wiremesh at the outlet of the expansion, Fig.~\ref{fig:fig3}(e), only the plane mode remains in the large tube.
This method can be used practically to optimally produce a plane mode in a guide where several other modes are propagating by using a source in the small guide.

\section{Conclusions}
Inserting a wiremesh into the waveguide parallel to its walls is an efficient method of modal attenuation and higher-order modes  filtering. For the critical location of the wiremesh $d_c$, varying the resistance of the wiremesh one can achieve the exceptional point of the first two higher order modes, where analogously to the Cremer criterion, the strongest attenuation is expected. This allows for  efficient plane wave selection in a waveguide with multiple propagating modes.

\section{Acknowledgments}
This work was supported by the French National Agency for Research (SelfiXs Project, ANR-18-CE92-0001). The authors have no conflicts of interest to declare. The data that support the findings of this study are available from the
corresponding author upon reasonable request.

\bibliography{main}
\end{document}